\documentclass[10pt]{article}
\usepackage{graphicx}

\usepackage{dcolumn}
\usepackage{bm}
\usepackage{lscape}
\usepackage{amsmath,amssymb,graphicx}
\usepackage{epsfig}
\usepackage{pstricks}
\usepackage{multirow}
\usepackage{booktabs}
\usepackage{axodraw4j}
\usepackage{subfigure}
\usepackage{cite}
\usepackage{rotating}
\usepackage{float}
\usepackage{soul}
\usepackage{comment}
\usepackage{textcomp}
\usepackage[english]{babel}
\usepackage{afterpage}
\usepackage{multicol} 

\usepackage{url}
\usepackage{picins}

\let\oldbibliography\thebibliography
\renewcommand{\thebibliography}[1]{\oldbibliography{#1}
\setlength{\itemsep}{0pt}} 

\usepackage{array}
\newcolumntype{L}[1]{>{\raggedright\let\newline\\\arraybackslash\hspace{0pt}}m{#1}}
\newcolumntype{C}[1]{>{\centering\let\newline\\\arraybackslash\hspace{0pt}}m{#1}}
\newcolumntype{R}[1]{>{\raggedleft\let\newline\\\arraybackslash\hspace{0pt}}m{#1}}

\RequirePackage{lineno}

\newcommand{\Z}{\mathbb Z_2}

\newcommand{\MET}{\emph{$E_{\rm T}^{\rm miss}$}}

\newcommand{\compresslist}{ 
\setlength{\itemsep}{1pt}
\setlength{\parskip}{0pt}
\setlength{\parsep}{0pt}
}

\def\Title#1{\begin{center} {\Large #1 } \end{center}}
\def\Author#1{\begin{center}{ \sc #1} \end{center}}
\def\Address#1{\begin{center}{ \it #1} \end{center}}

\newcommand\pubblock{\rightline{\begin{tabular}{l} Proceedings of the Fifth Annual LHCP\\ \pubnumber\\
         \pubdate  \end{tabular}}}

\newenvironment{Abstract}{\begin{quotation} \begin{center} 
             \large ABSTRACT \end{center}\bigskip 
      \begin{center}\begin{large}}{\end{large}\end{center} \end{quotation}}

\newenvironment{Presented}{\begin{quotation} \begin{center} 
             PRESENTED AT\end{center}\bigskip 
      \begin{center}\begin{large}}{\end{large}\end{center} \end{quotation}}

\def\beq{\begin{equation}}
\def\eeq#1{\label{#1}\end{equation}}
\def\eeqn{\end{equation}}

\def\beqa{\begin{eqnarray}}
\def\eeqa#1{\label{#1}\end{eqnarray}}
\def\eeqan{\end{eqnarray}}

\let\bar=\overbar

\def\L{{\cal L}}

\def\Dslash{\not{\hbox{\kern-4pt $D$}}}
\def\dslash{\not{\hbox{\kern-2pt $\del$}}}

\def\msb{{\bar{\ssstyle M \kern -1pt S}}}

\usepackage{color}

\newcommand\pubnumber{ }

\newcommand\pubdate{\today}

\def\affiliation{\small $^1$\textit{School of Physics and Astronomy, University of Southampton, Southampton SO17 1BJ, UK}\\[0pt]
\vspace*{0.1cm} $^2$\textit{Particle Physics Department, Rutherford Appleton Laboratory, Chilton, Didcot, Oxon OX11 0QX, UK}\\[0pt]
\vspace*{0.1cm} $^3$\textit{Physics Department, CERN, CH-1211, Geneva 23, Switzerland}\\[0pt]
\vspace*{0.1cm} $^4$\textit{Dipartimento di Fisica, Universit\`a di Genova and INFN, Sezione di Genova,\\via Dodecaneso 33, 16146 Genova, Italy}}

\begin{document}

\def\thefootnote{\fnsymbol{footnote}}
\setcounter{footnote}{1}

\large
\begin{titlepage}
\pubblock

\vfill
\Title{Extra Quarks Decaying to Dark Matter \\[0.25cm] Beyond the Narrow Width Approximation}
\vfill

\Author{Hugo Prager\footnote{Speaker.}$^{,1,2}$, Stefano Moretti$^{1,2,3}$, Dermot O'Brien$^{1,2}$, Luca Panizzi$^{1,2,4}$}

\Address{\affiliation}

\vfill
\begin{Abstract}

We explore the effects induced by a finite width in processes of pair production of a heavy top-quark partner and its subsequent decay into a bosonic Dark Matter (DM) candidate -- either scalar or vector -- and a  SM up-type quark at the Large Hadron Collider (LHC). We discuss the configurations of masses, widths and couplings where this phenomenology can be important in a simple model with just one such objects.
Finally, we emphasise the correct definition of signal and background to be adopted as well as stress the importance of new dedicated experimental searches.   

\end{Abstract}
\vfill

\begin{Presented}
The Fifth Annual Conference\\
 on Large Hadron Collider Physics \\
Shanghai Jiao Tong University, Shanghai, China\\ 
May 15-20, 2017
\end{Presented}
\vfill
\end{titlepage}
\def\thefootnote{\fnsymbol{footnote}}
\setcounter{footnote}{0}
%

\normalsize

\vspace{-4mm}

\section{Introduction}

\vspace{-2mm}

Understanding the nature of potential eXtra Quarks (XQs)  and evident DM  may  possibly mean to study the 
two sides of the same coin, as these states appear together in a variety of Beyond the SM (BSM) scenarios, e.g., in Universal Extra Dimensions \cite{Antoniadis:1990ew,Cacciapaglia:2009pa}), where indeed the DM candidate (scalar or vector)  can be produced via the decay of an XQ. In this kind of scenarios, we get signatures with
significant  missing  transverse energy
($\MET$), hence  similar to the case of Supersymmetry (SUSY) with conserved $R$-parity, an aspect which makes it possible to interpret SUSY results in the Narrow Width Approximation (NWA) in terms of limits on (narrow) XQs \cite{Cacciapaglia:2013wha,Kraml:2016eti}. The main benefit of this is of course that this procedure, based on the NWA assumption, makes the approach model independent, as the Branching Ratio (BR) of the XQ into DM (and a standard quark)
is the only unknown BSM parameter, apart from the XQ mass, since the dominant production channel of XQs at the LHC
is QCD induced pair production. However, this approach is rather limiting in terms of the XQ parameters that can be probed, 
as the XQ can have a large decay width. In fact, here we want to evaluate the effects of \emph{a large width} of the XQ 
in the determination of its cross-section and, hence, in the (re)interpretation of bounds from experimental searches. Recall that, in these circumstances, i.e., when the XQ is off-shell, in order to preserve gauge invariance, additional diagrams should be included in the definition of the signal, as we shall do below.  These will include topologies with one XQ only. 
As immediate consequence of this, we notice that such an approach becomes necessarily model dependent.  
It is the purpose of this writeup to discuss the salient features of this phenomenology, borrowing results from   \cite{Moretti:2017qby}. 

\vspace{-3mm}

\paragraph{Lagrangian terms}

The interaction between a \textit{singlet} DM (scalar or vector) and the XQ for a coupling with third generation SM quark is described by the following lagrangian terms\footnote{The interaction terms for a coupling with first and second generation quark are completely analogous.} 
\vspace{-2mm}
\begin{eqnarray}
\L^S_1 &=& 
\left[
\lambda_{11}^t \bar{T} P_R \; t + 
\lambda_{11}^b \bar{B} P_R \; b +
\lambda_{21} \; \overline\Psi_{1/6} P_L {t \choose b} 
\right] 
S^0_{DM} \notag
\label{eq:LagSingletDMS}
\\
\L^V_1 &=& 
\left[
g_{11}^t \bar{T} \gamma_\mu P_R \; t + 
g_{11}^b \bar{B} \gamma_\mu P_R \; b + 
g_{21}  \; \overline\Psi_{1/6} \gamma_\mu P_L {t \choose b} 
\right] 
V^{0\mu}_{DM} \notag 
\label{eq:LagSingletDMV}
\end{eqnarray}
where $T$ and $B$ represent XQ singlets and $\Psi_{1/6}=(T \ B)^T$ a XQ doublet, while $S^0_{DM}$ and $V^0_{DM}$ represent scalar and vector DM. All the new particles are odd under a $\Z$ symmetry which is needed to make the DM stable under the assumption that all the SM states are even under the same symmetry. Notice that depending on the values of the coupling constants the XQ can be either vector-like or chiral\cite{Kraml:2016eti}.

Our results have been obtained considering the \emph{vector-like $T$} component of a $(T \ B)^T$ doublet, although we checked that singlet and chiral $T$ XQs produce analogous results.

\vspace{-3mm}

\paragraph{Observables and conventions}

Two different processes leading to the same final state $DM \; DM \ q \; \bar{q}$ are considered: 
\vspace{-2mm}

\begin{itemize} \compresslist
\item \textit{QCD pair production and on-shell decay}: in the NWA regime the cross-section is given by $\sigma_X (M_Q) \equiv \sigma_{2 \to 2}^{QCD}~BR(Q)~BR(\bar Q)$ and this process is usually considered in experimental searches for XQ pair production,
\item \textit{Full signal}: all the topologies containing at least one XQ are taken into account, including some which are missing in the NWA, and the cross-section will be labelled by $\sigma_S (M_Q, M_{\rm DM}, \Gamma_Q)$,
\end{itemize}
\vspace{-2mm}
and we study the ratio $(\sigma_S - \sigma_X)/\sigma_X$ to provide a quantitative measure of the difference between the two processes.

\vspace{-3mm}

\paragraph{Analysis tools and experimental searches\label{sec:Monte Carlo}} 

We perform our study by scanning over three free parameters: the masses $M_T$, $M_{\rm DM}$ and the width $\Gamma_T$.
We consider scenarios where the DM state has masses $M_{\rm DM}$ = 10 GeV and 500 GeV\footnote{The case of DM with mass of 1 TeV has also been considered in \cite{Moretti:2017qby} but it will not be presented here.} and where the XQ has mass $M_T > M_{\rm DM} + m_q$, with $q \in \{u,c,t\}$ (such that its on-shell decay is kinematically allowed) up to $M_T^{\rm max}$ = 2500 GeV, which is the maximal value to obtain at least 1 signal event for $T$ pair production at LHC@13TeV with 100/fb integrated luminosity. The $T$ width ranges from $\Gamma_T / M_T$ $\simeq$ 0\% (NWA) to 40\% of the $T$ mass.

Our numerical results at partonic level are obtained using {\sc MadGraph5} \cite{Alwall:2011uj,Alwall:2014hca} and a model we implemented in {\sc Feynrules} \cite{Alloul:2013bka}. Events are simulated using the PDF set {\sc cteq6l1}~\cite{Pumplin:2002vw} and then passed to {\sc Pythia}\,8~\cite{Sjostrand:2007gs,Sjostrand:2006za}, which takes care of the hadronisation and parton showering.

To analyse our results and perform a recasting with a set of 13 TeV analyses considering final states compatible with our scenarios, we employ {\sc CheckMATE 2}~\cite{Dercks:2016npn}, which uses {\sc Delphes\,3}~\cite{deFavereau:2013fsa} for the emulation of detector effects. We consider all the ATLAS and CMS analyses at 13 TeV available within the CheckMATE database, the most relevant ones in our case being ATLAS 1604.07773 \cite{Aaboud:2016tnv}, ATLAS 1605.03814 \cite{Aaboud:2016zdn}, ATLAS-CONF-2016-050 \cite{ATLAS-CONF-2016-050}.

\vspace{-2mm}

\section{Extra $T$ quark interacting with DM and the SM top quark}
\label{sec:Tt}  

The possible decay channels for a XQ coupling to third generation SM quarks are $t \bar t + \{S^0_{DM} S^0_{DM}, V^0_{DM} V^0_{DM}\}$, {i.e.} $t \bar t + \MET$. 

\vspace{-2mm}

\paragraph{Large width effects at parton level}

We show in Fig.\ref{fig:SXthird} the ratio $(\sigma_S - \sigma_X)/\sigma_X$ for an LHC energy of 13 TeV for two representative cases.

\piccaption{\label{fig:SXthird} Relative difference between full signal and NWA cross-sections. Left: vector DM of mass 10 GeV; right: scalar DM of mass 500 GeV.}
\parpic{
\epsfig{file=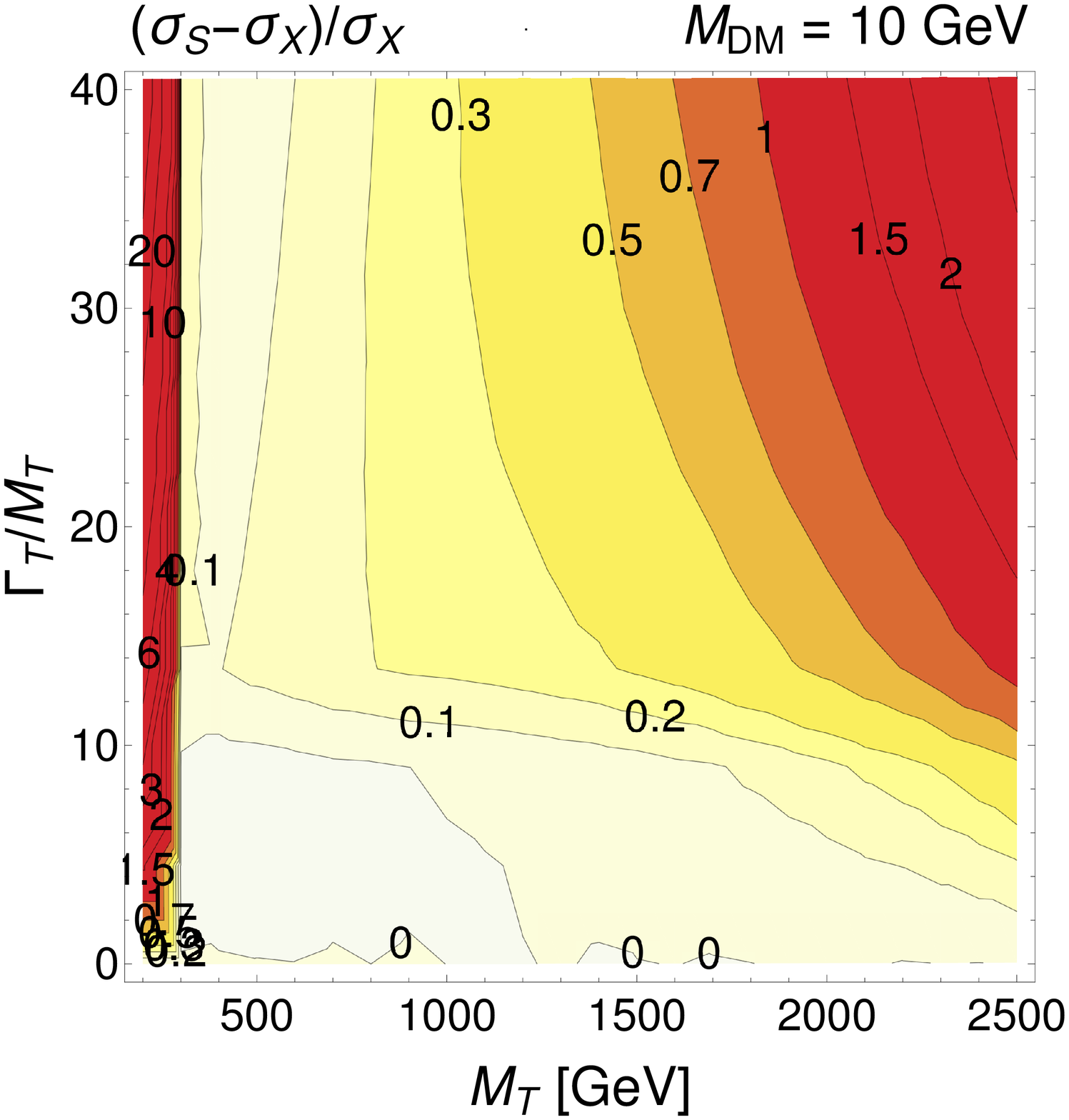, width=.26\textwidth}
\epsfig{file=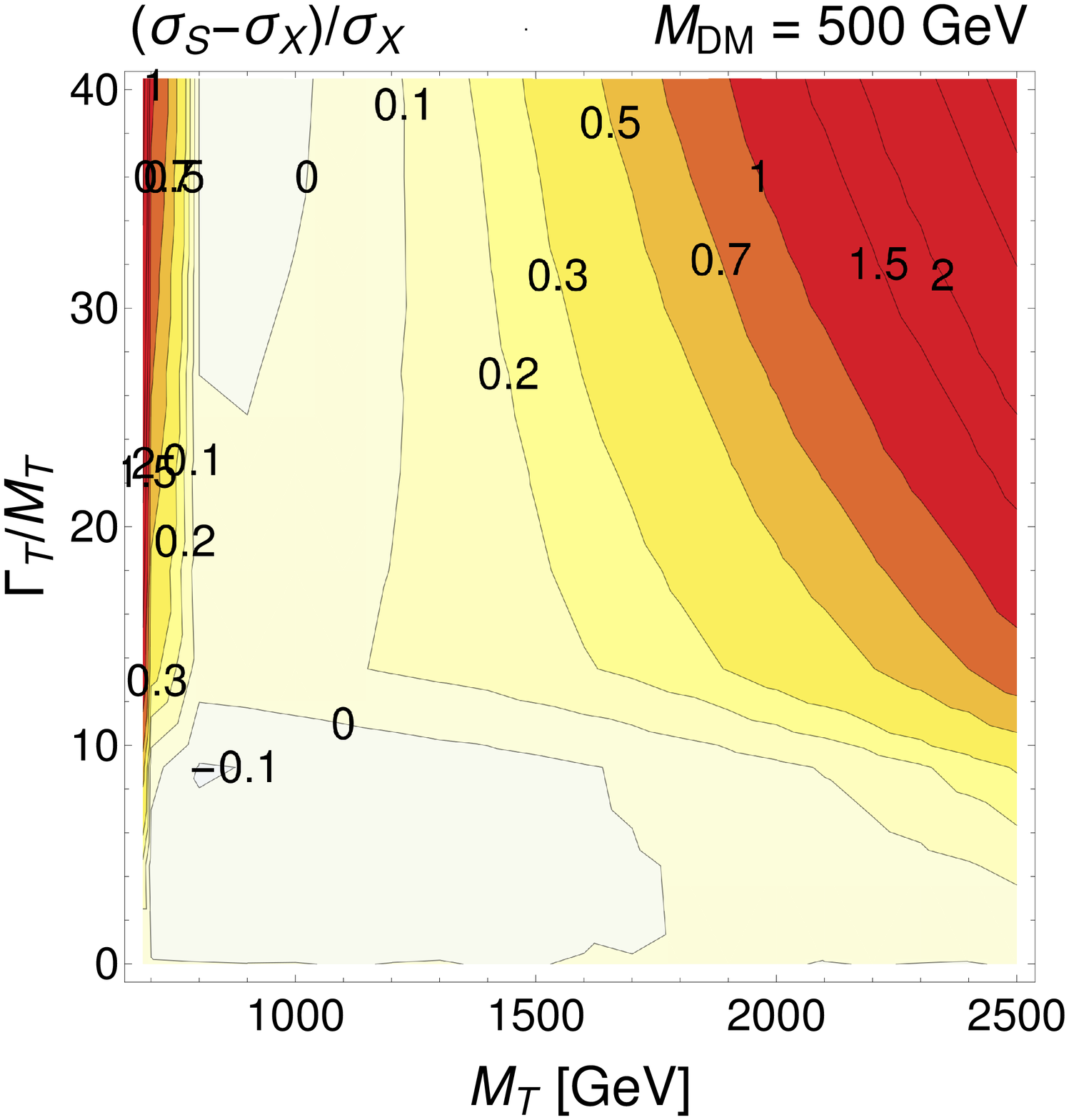, width=.26\textwidth}
}

As expected, the off-shell contributions are \emph{negligible} in the NWA and become more and more relevant as $\Gamma_T$ increases. The increase is especially fast when the $T$ mass approaches the kinematic limit, and this can be explained by a non-trivial combination of factors: as the gap between the masses decreases, a larger $T$ width opens a larger phase space for $T$ decay. Moreover, in some regions $\sigma_X$ becomes similar to $\sigma_S$ even for large values of the width; nevertheless, the differential distributions of the signal in the NWA and finite width are different and therefore, it is not possible to describe the process through the NWA approximation.

\vspace{-2mm}

\paragraph{Large width effects at detector level}

We show in Fig.\ref{fig:Exclusion3} the results obtained with the searches implemented in {\sc CheckMATE} for a DM particle of mass 10 GeV and 500 GeV.

\piccaptiontopside\piccaption{\label{fig:Exclusion3}{\sc CheckMATE} results for a DM particle of mass 10 GeV and 500 GeV. We show in black (grey) the exclusion line for the scalar (vector) DM scenario.}
\parpic[right]{
\epsfig{file=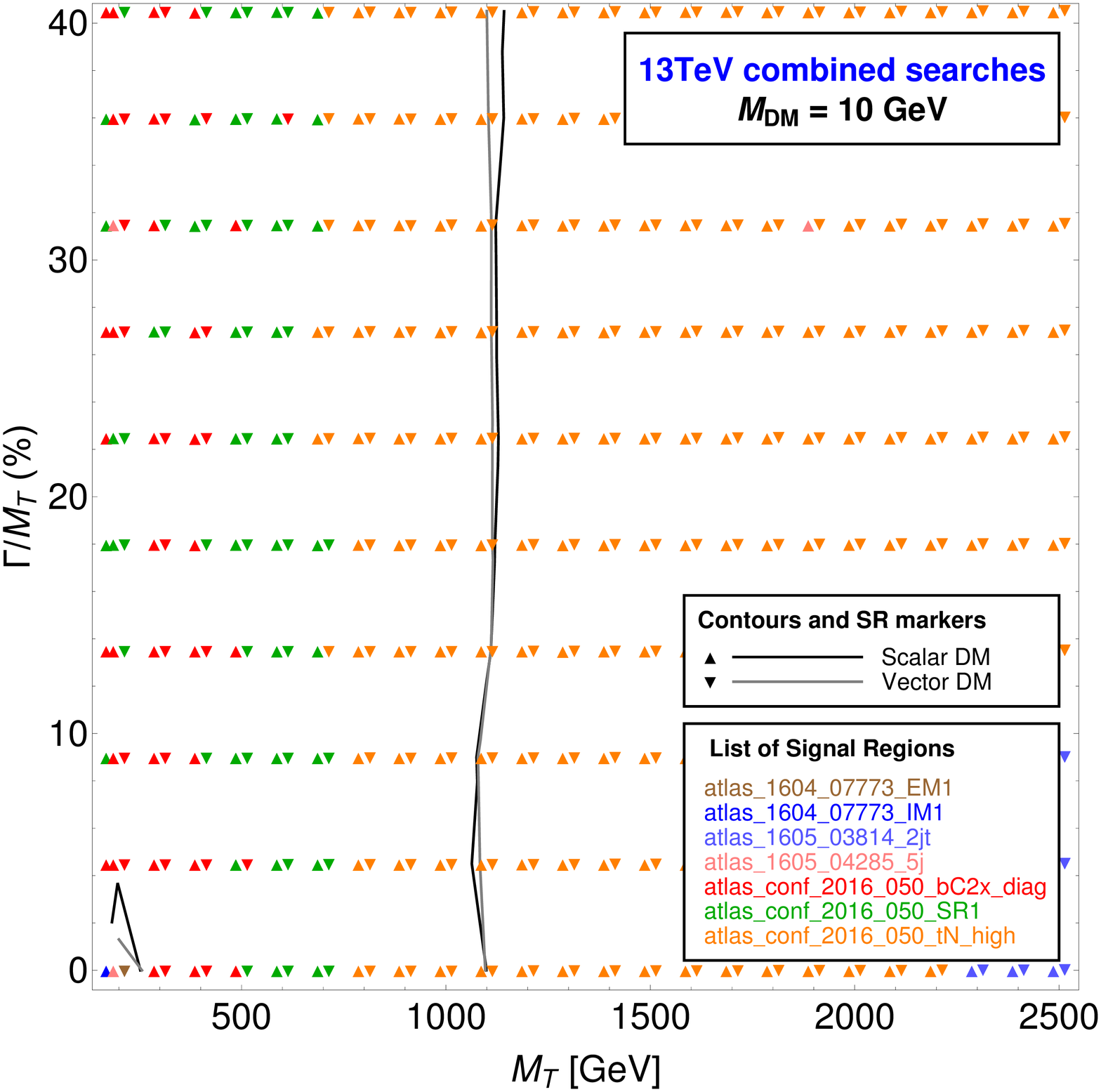, width=.26\textwidth} 
\epsfig{file=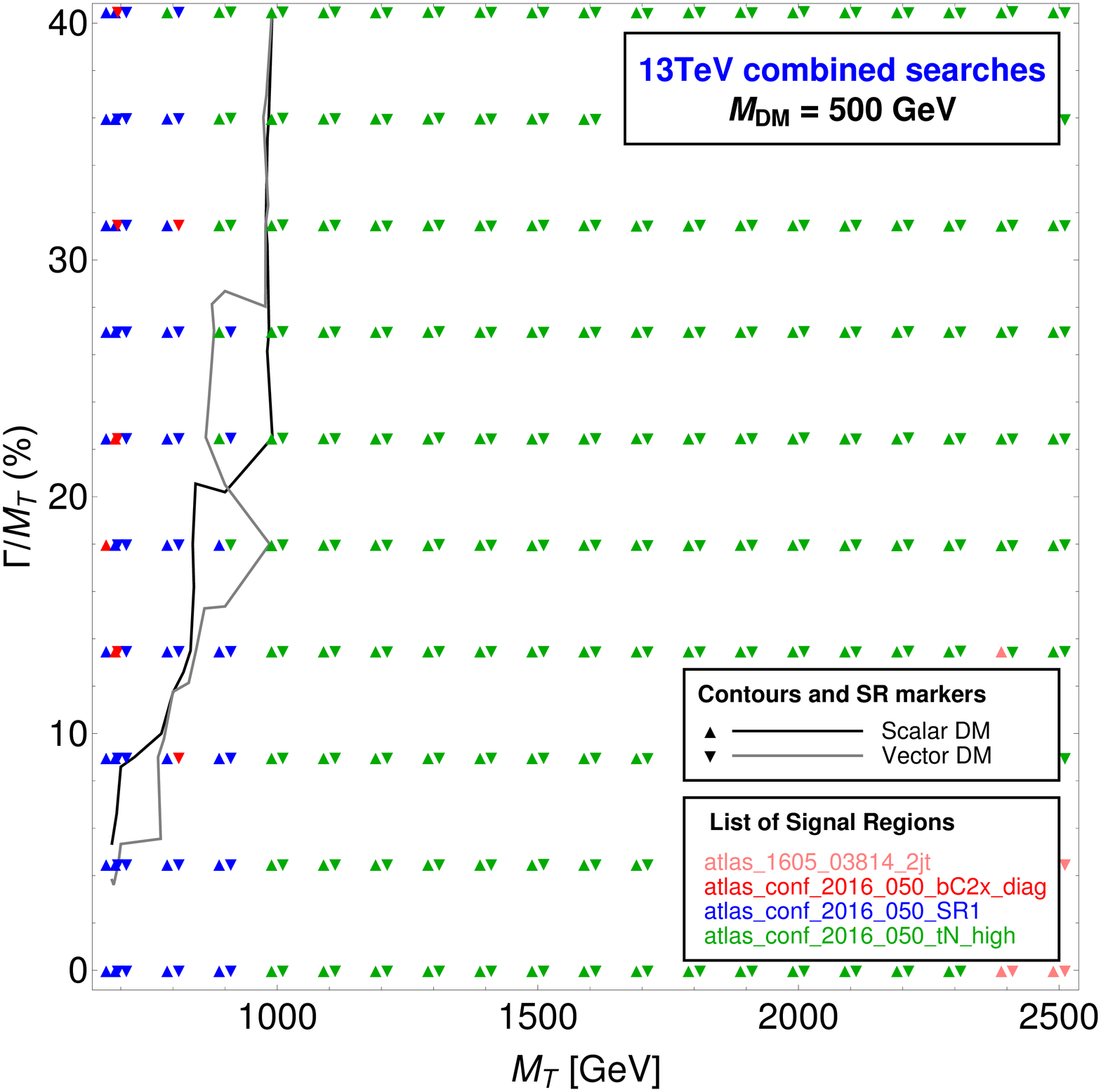, width=.26\textwidth} 
}

The width dependence of the bounds is very weak in both cases, and the numerical values of the bounds for scalar and vector DM are similar, despite the fact that the signal cross-sections are different. The weak dependence of the bounds on the width is determined by a combination of cross-section and selection efficiencies: the cuts on $p_T^{\rm jet}$ and $\MET$ have a strong effect on the signal and result in smaller efficiencies when $\Gamma_T$ increases. 

\piccaptiontopside\piccaption{\label{fig:sigmaandeffs3} Left: Signal cross-section; right: efficiency of the most relevant signal region.}
\parpic{
\epsfig{file=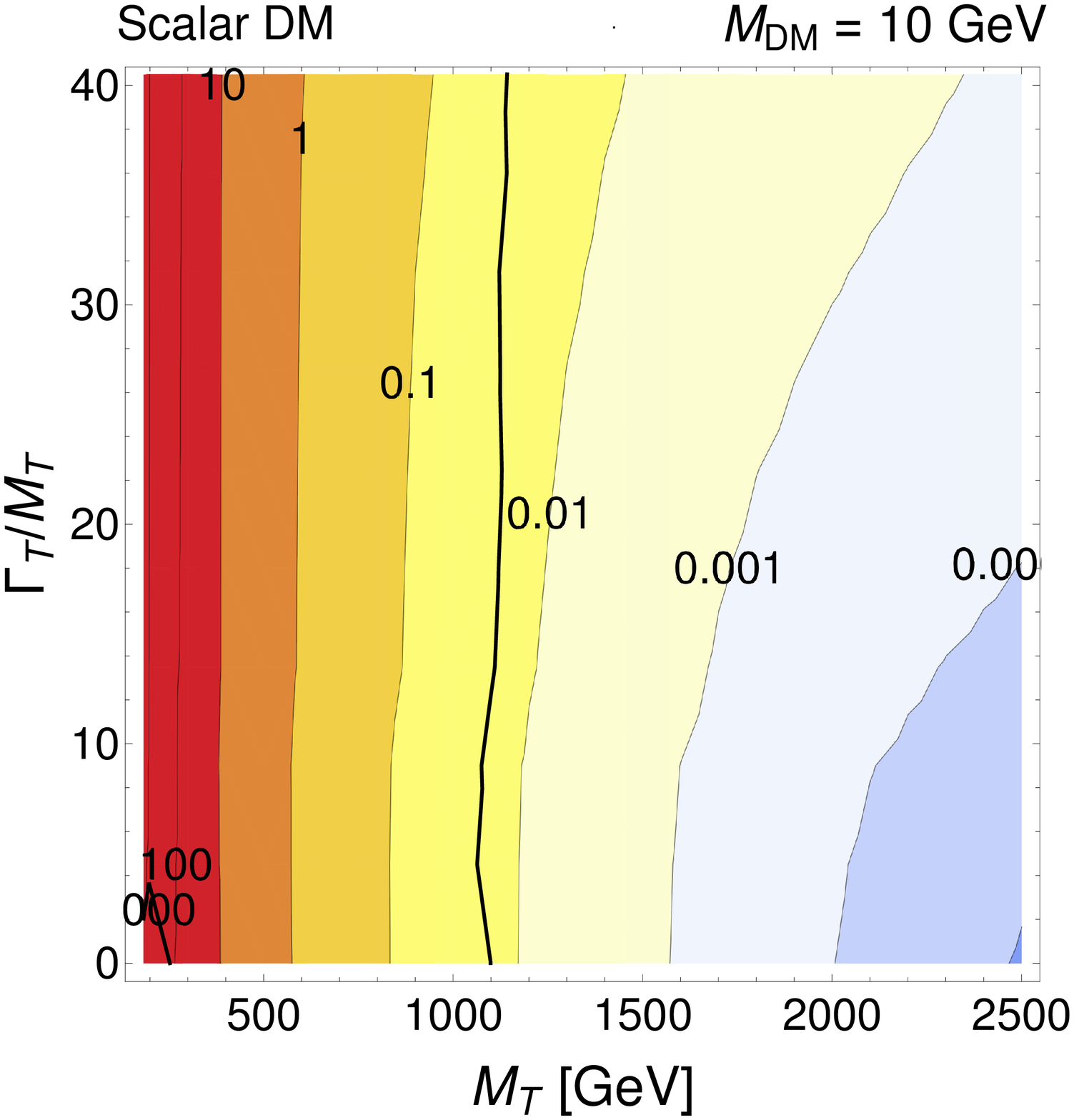, width=.26\textwidth} 
\epsfig{file=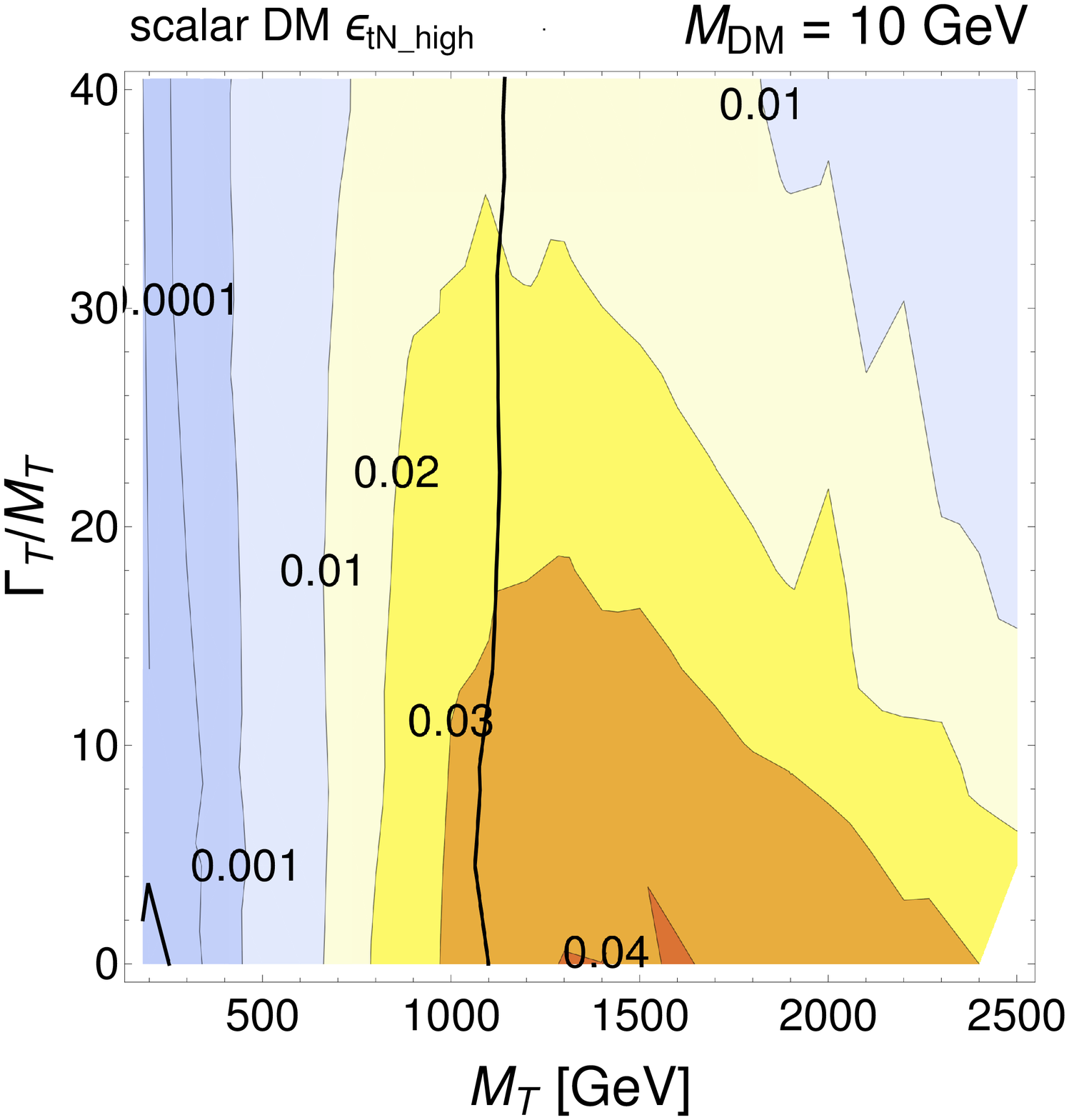, width=.26\textwidth} 
}

We show in Fig.\ref{fig:sigmaandeffs3} the signal cross-section and the efficiency for the most relevant signal region together with the exclusion line for a scalar DM of mass 10 GeV. The plot clearly shows that the increase of cross-section is compensated by the decrease of efficiency. Therefore the NWA hypothesis made for experimental searches \emph{does not overestimate the bound}, even if the $T$ has a large width. 

\vspace{-2mm}

\section{Extra $T$ quark interacting with DM and the SM up quark}
\label{sec:Tu}

\parpic[right]{
\begin{picture}(100,60)(-10,-5)
\SetWidth{1}
\Line[arrow](0,0)(25,0)
\Text(-2,0)[rc]{\large $u$}
\Line[arrow](25,45)(0,45)
\Text(-2,45)[rc]{\large $\bar u$}
\Photon(20,0)(70,0){3}{9}
\Line[dash](25,0)(70,0)
\Text(72,0)[lc]{\large DM}
\SetColor{Red}\SetWidth{1.5}
\Line[arrow](25,0)(25,22.5)
\Text(21,12)[rc]{\large \Red{$T$}}
\Line[arrow](25,22.5)(25,45)
\Text(21,32)[rc]{\large \Red{$T$}}
\SetColor{Black}\SetWidth{1}
\Photon(25,45)(70,45){3}{9}
\Line[dash](25,45)(70,45)
\Text(72,45)[lc]{\large DM}
\Gluon(25,22.5)(50,22.5){3}{6}
\Line[arrow](50,22.5)(71,32)
\Text(72,32)[lc]{\large $u$}
\Line[arrow](70,10)(50,22.5)
\Text(72,10)[lc]{\large $\bar u$}
\end{picture}
}

We now study the case of XQs coupling a DM candidate and first generation SM quarks. The possible final states are $S^0_{DM}u \; S^0_{DM}\bar u$ and $\ V^0_{DM}u \; V^0_{DM}\bar u$, {i.e.}  $j j + \MET$.

In this case new topologies containing collinear divergences (due to the gluon splitting) are present, such as the one showed on the right.  
These diagrams affect the signal but not the QCD pair production, therefore giving rise to a large increase of the ratio $(\sigma_S - \sigma_X) / \sigma_X$. For this reason we use logarithmic plots in the following.

\vspace{-2mm}

\paragraph{Large width effects at parton level}

We show in Fig.\ref{fig:SXfirst} the relative difference between full signal and NWA cross-sections for representative scenarios.

\piccaption{\label{fig:SXfirst} Relative difference between full signal and NWA cross-sections. Left: vector DM of mass 10 GeV; right: scalar DM of mass 500 GeV.}
\parpic{
\epsfig{file=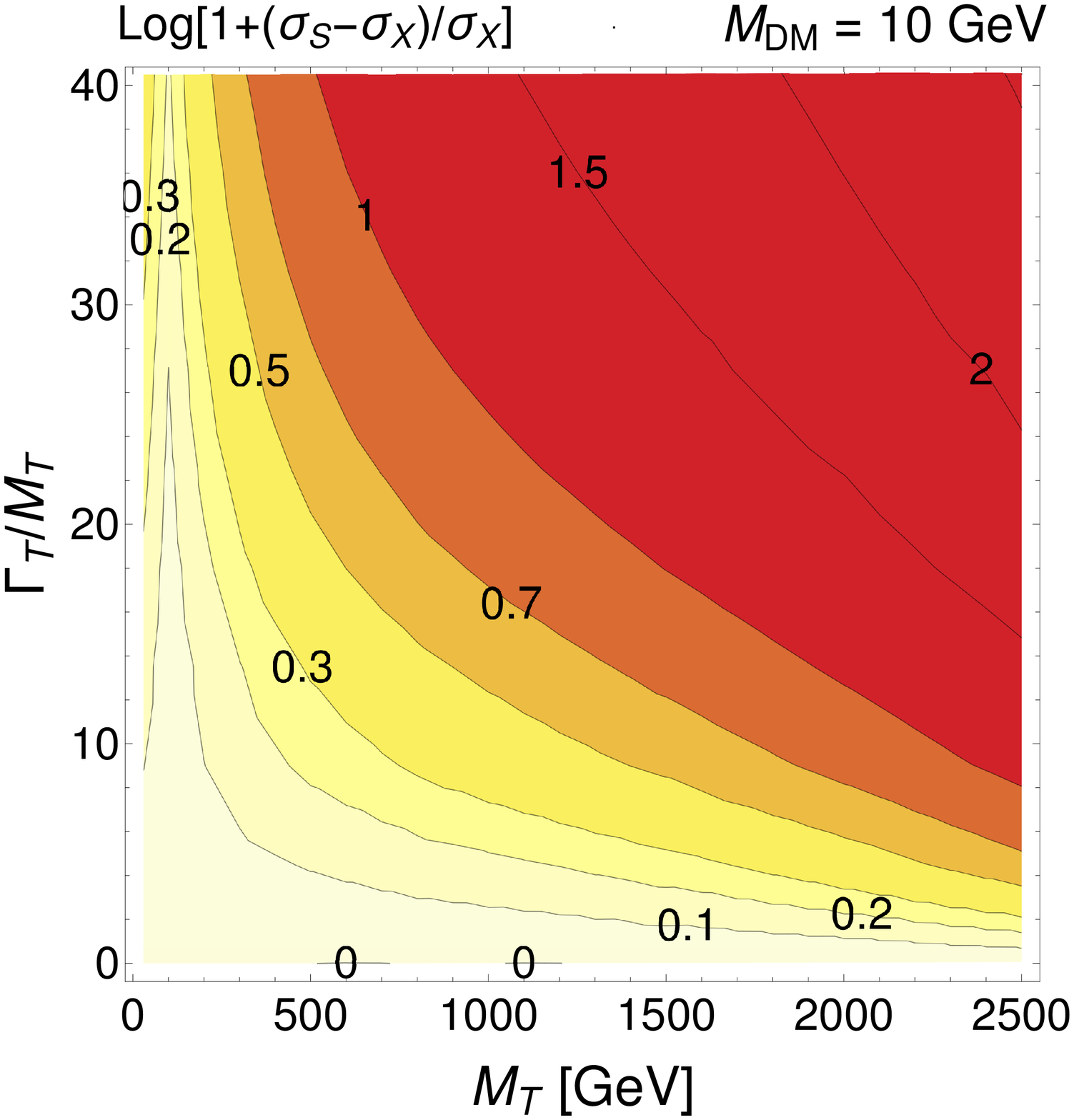, width=.26\textwidth} 
\epsfig{file=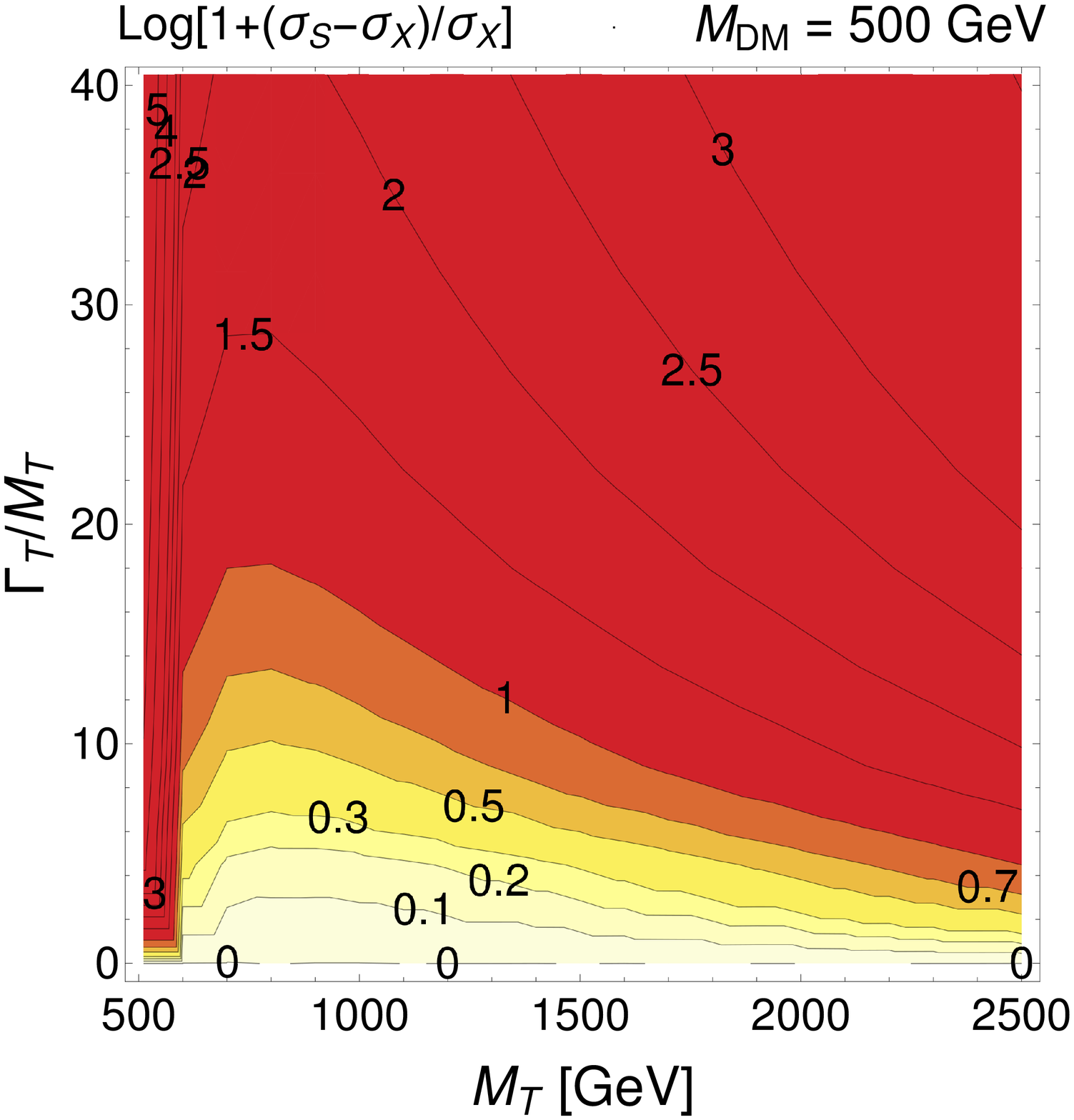, width=.26\textwidth} 
}
The ratio is much larger than for third generation coupling and correctly approaches zero in the NWA. There are no regions where cancellations make $\sigma_S$ similar to $\sigma_X$ for large width, but the cross-section ratio still decreases in regions that are very similar to the ones observed for 3rd generation. This means that even if a cancellation of opposite effects still takes place, the effect of the additional diagrams is dominant. \\ 

\vspace{-2mm}

\paragraph{Large width effects at detector level}
In Fig.\ref{fig:detector1} we observe a sizable width dependence of the bounds, which are still similar for scalar and vector DM in the NWA, as expected. Furthermore, it is possible to distinguish scalar and vector DM.

The shape of the bounds is again fetermined by a combination of cross-section and efficiencies contributions. The cross-section effect is dominant in this case: for both scalar and vector DM, the bounds \emph{track the different behaviours in the scaling of $\sigma_S$}. In Fig.\ref{fig:detector1} we have shown examples with one of the two DM spins.

The NWA hypothesis made by experimentalist may \emph{underestimate the bounds} if the $T$ width is large, as the bounds may reach values larger than 2 TeV. 

\hspace{-5mm}\begin{minipage}{0.48\textwidth}
\centering
\epsfig{file=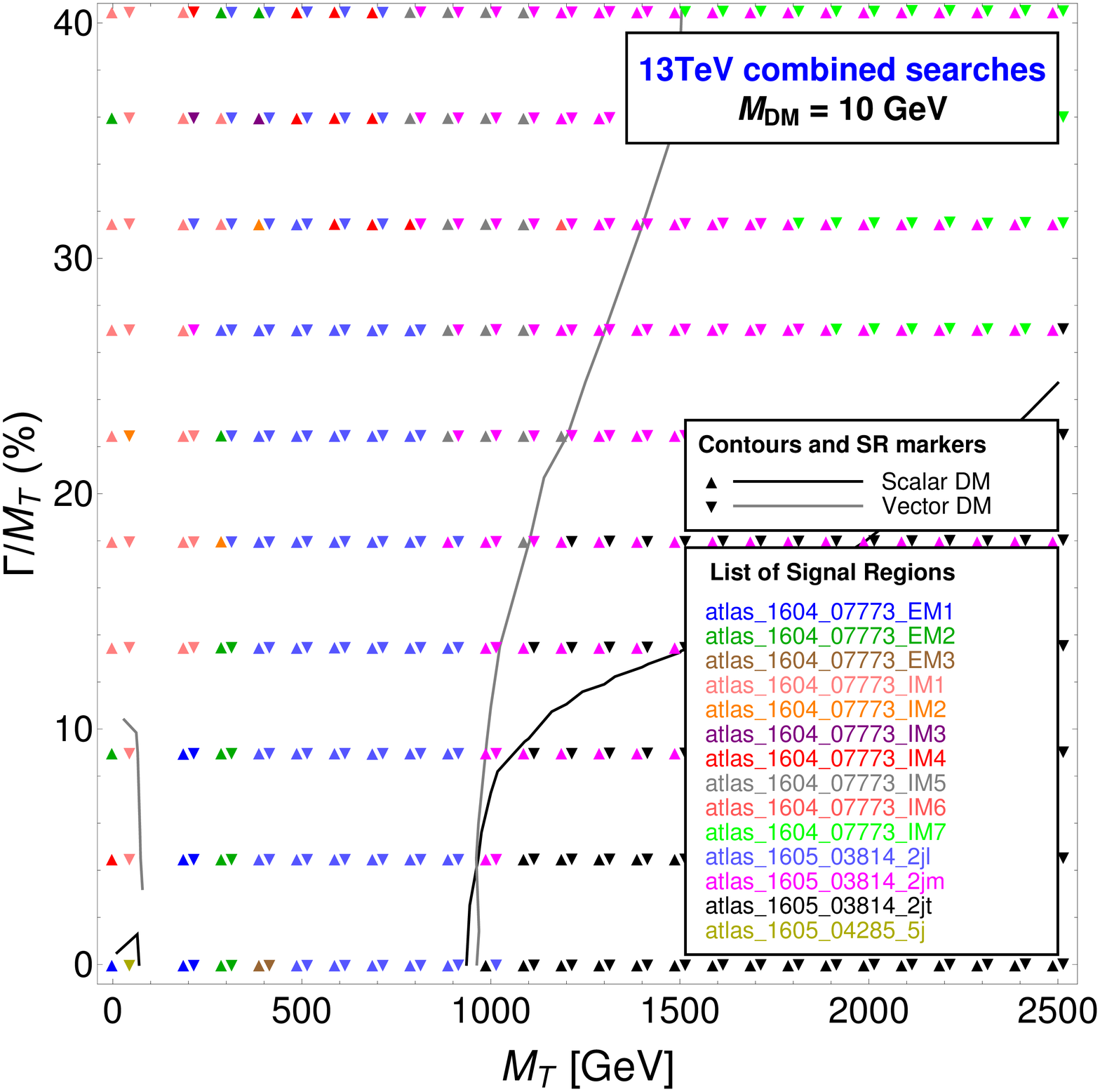, width=.49\textwidth} 
\epsfig{file=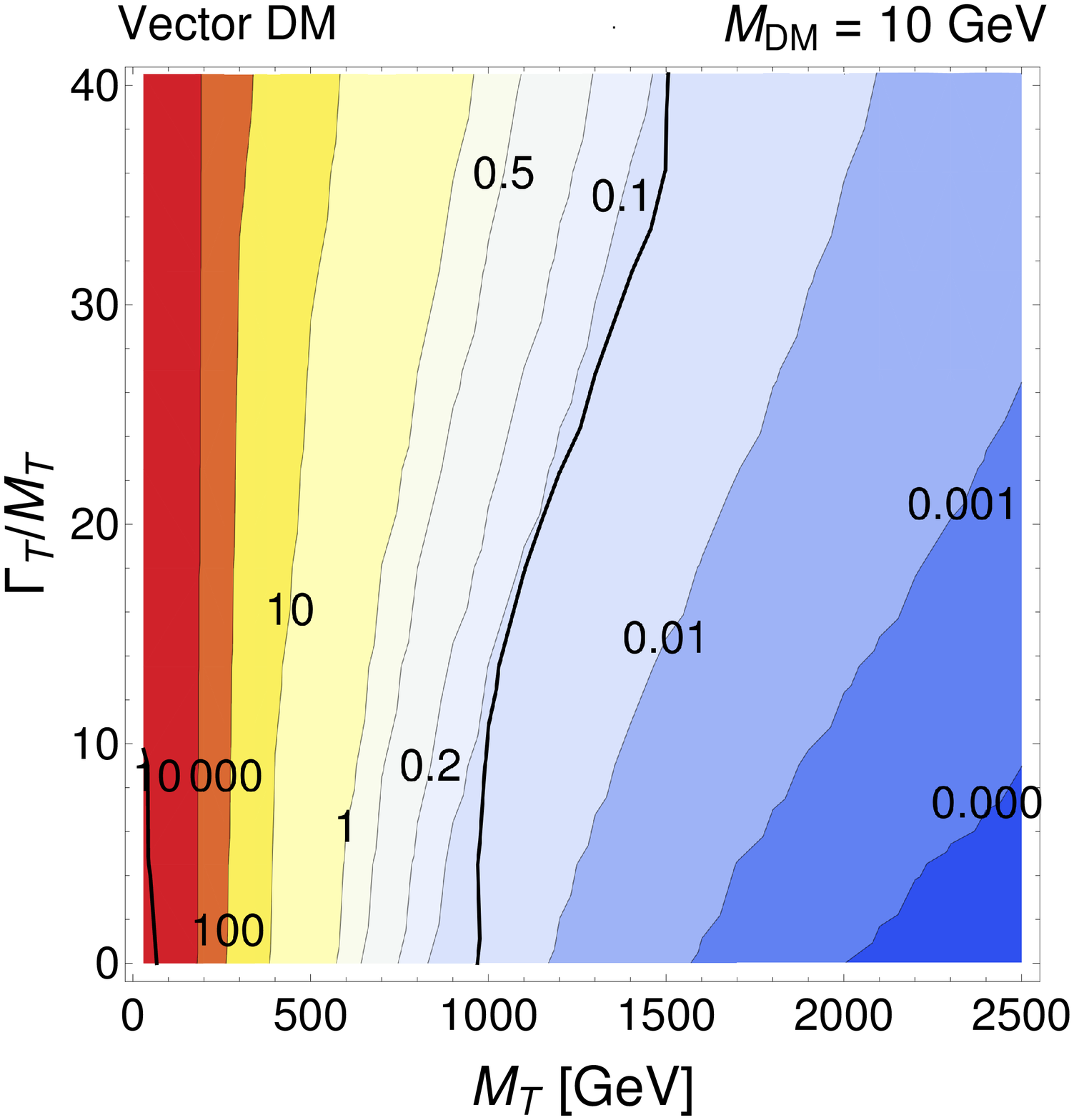, width=.49\textwidth}
\end{minipage}\hfill
\begin{minipage}{0.48\textwidth}
\centering
\epsfig{file=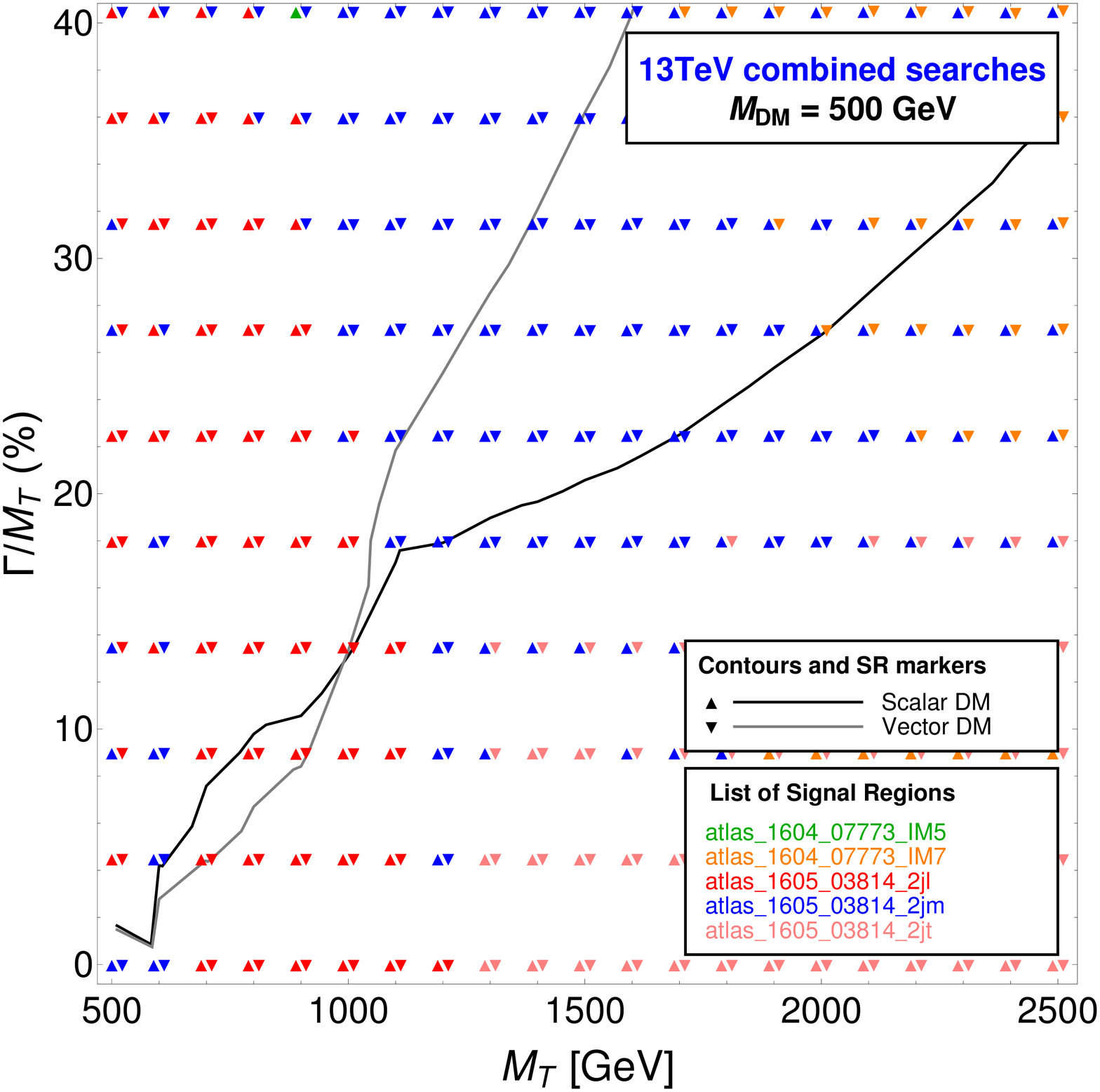, width=.49\textwidth} 
\epsfig{file=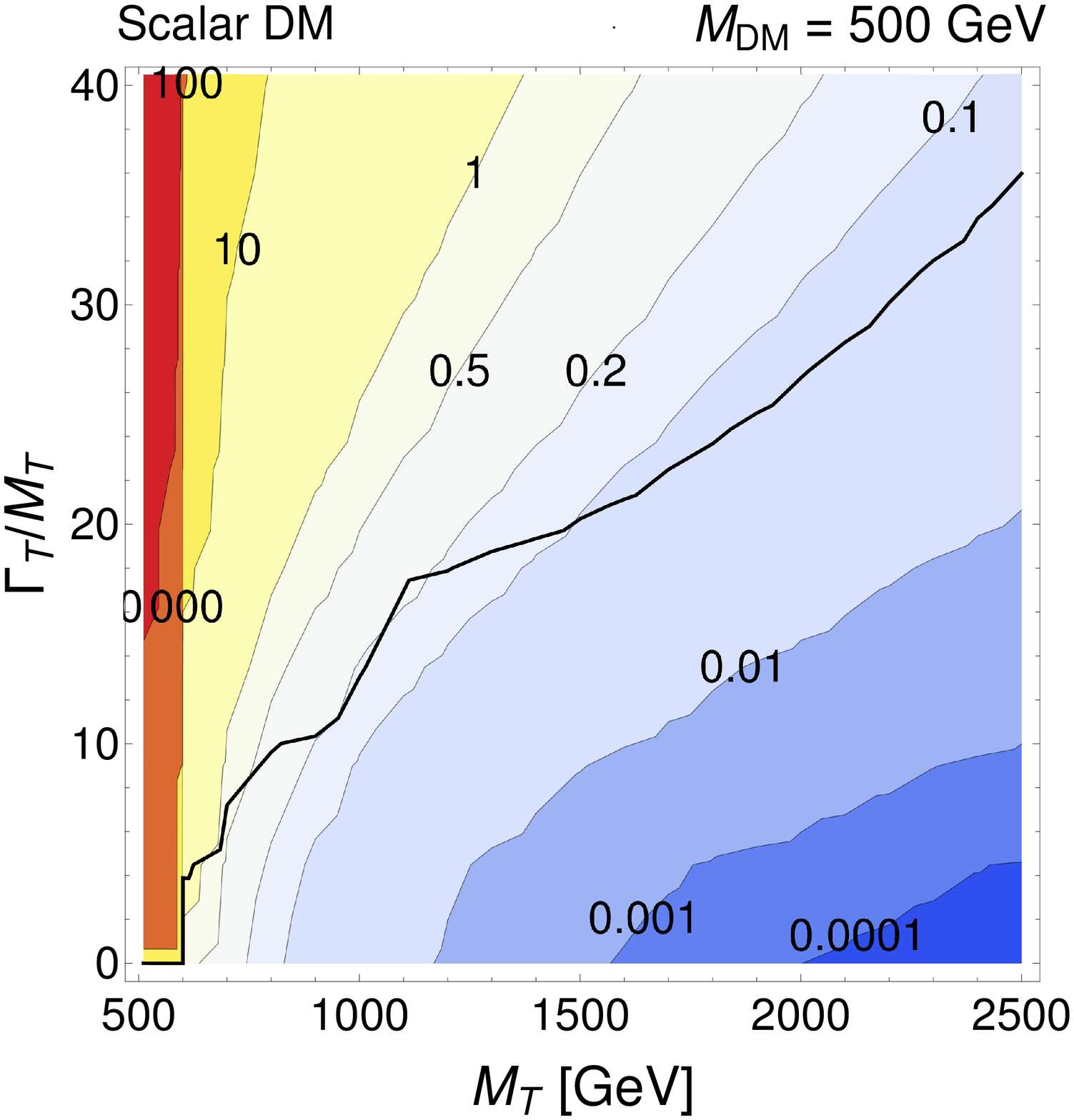, width=.49\textwidth} 
\end{minipage}
\vspace{-2mm}
\caption{\label{fig:detector1} Exclusion bounds and signal cross-sections for DM with masses 10GeV and 500GeV.}

\vspace{-2mm}

\section{Exclusion limits in the $M_T-M_{DM}$ plane}
\label{sec:BellPlots}

\begin{figure}[h]
\centering
\epsfig{file=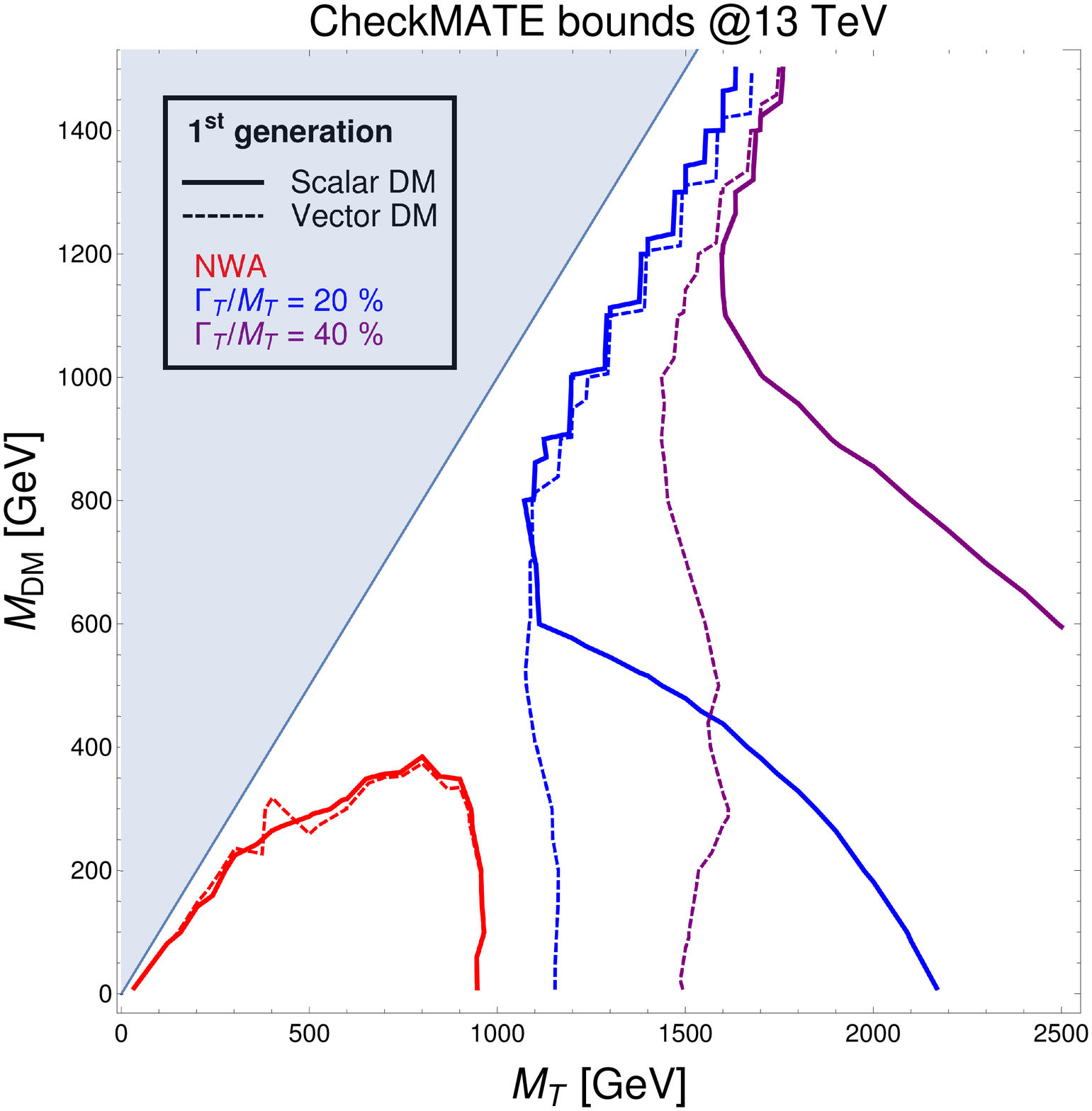, width=.32\textwidth} 
\epsfig{file=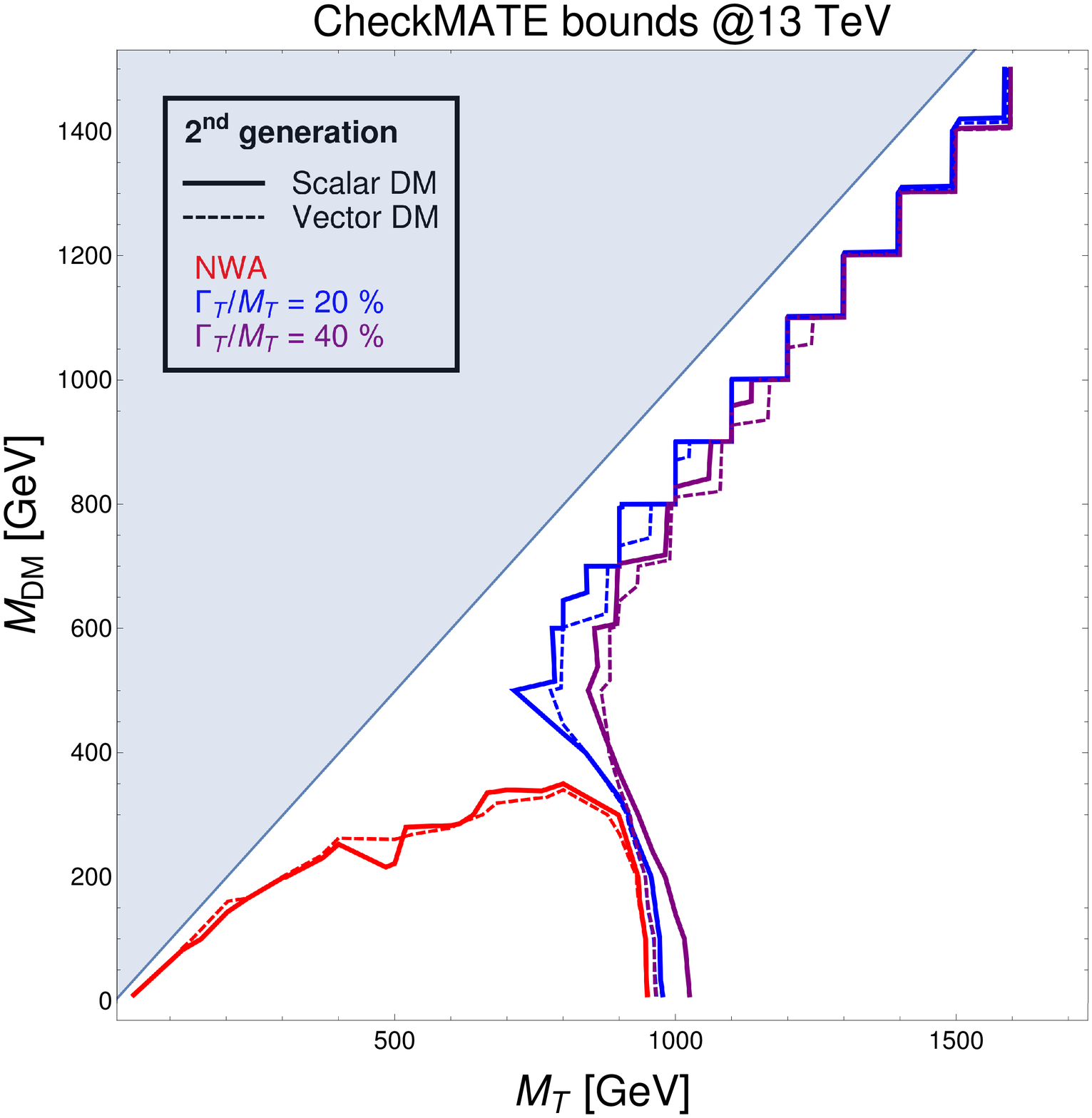, width=.32\textwidth} 
\epsfig{file=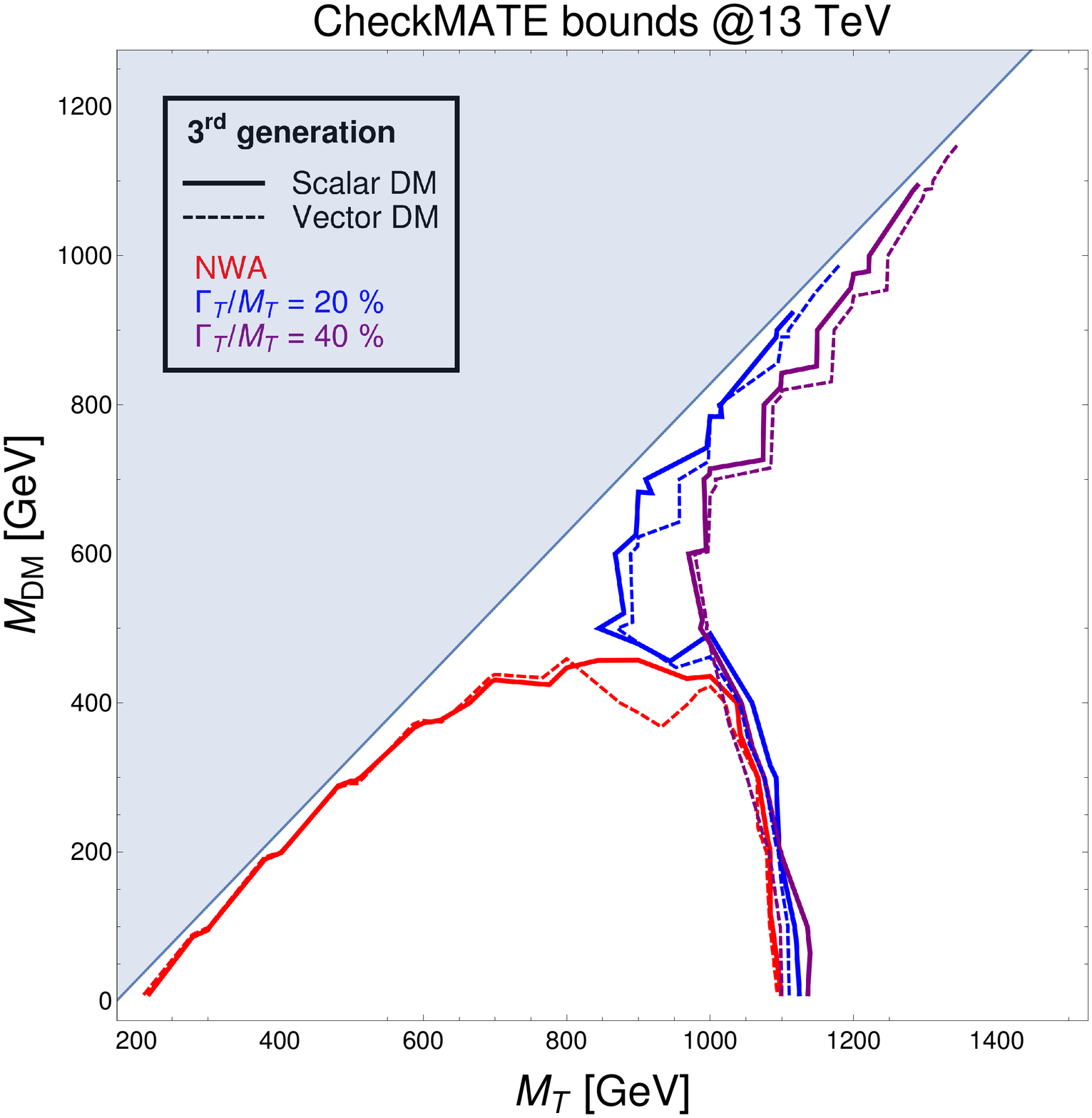, width=.32\textwidth} 
\vspace{-2mm}
\caption{\label{fig:bellplots}Exclusion limits in the $(M_T, M_{\rm DM})$ plane.}
\end{figure}

The scenarios we are considering are described by three parameters: the masses of the $T$ and of the DM and the width of the $T$. The 2$\sigma$ exclusion bound identifies a 3D surface in the space defined by such parameters and therefore it is instructive to study the projections of the surface on the plane identified by the masses of $T$ and DM for different values of the $\Gamma_T/M_T$ ratio. This allows us to directly compare bounds on $T$ and bosonic DM with analogous results in other models, such as supersymmetry. Exclusion limits for SUSY searches of stops are often presented in the $(M_{\tilde{t}}, M_{\chi_0})$ plane, where $\chi_0$ is the neutralino. Fig.\ref{fig:bellplots} shows the bounds in the $(M_T, M_{\rm DM})$ plane for specific values of $\Gamma_T / M_T$: the NWA case, 20\% and 40\%. Results for a $T$ quark coupling to DM and the charm quark are included in this figure.

The assumptions about the couplings of $T$ with SM quarks strongly affect the qualitative behaviours of the exclusion limits. In the NWA the bounds exhibit the \emph{same shape than the SUSY ones} whereas if the $T$ has a large width the region where \emph{the $T$ is almost degenerate with the DM is also excluded}, even if the masses are large, above the TeV. The \emph{dependence on the width} is stronger when the $T$ couples with the SM up quark, and in this case it is also possible to distinguish scalar from vector DM in the large width regime.

In all cases the bounds in the NWA are less stringent than those in the large width regime. The qualitative difference in the width dependence of the bounds, depending on the different assumptions about the $T$ couplings, is however remarkable. The design of new experimental searches could definitely take into account these effects to try achieve not only the discovery of new signals in channels with \MET\ but also a characterisation of the discovered signal.

\vspace{-2mm}


\section{Conclusions}
\label{sec:conclusions}

\vspace{-2mm}

In conclusion, results from model-dependent LHC searches for XQs decaying to DM (spin 0 or 1) and a  SM quark (which we have taken here to be both of $u$-type) do not account for effects induced by either the large XQ width,  additional XQ topologies (required by gauge invariance) or both, yet, these can be very large even in a simplified model with only one XQ, as we shown here. Therefore, in the attempt to make use of existing experimental results in such circumstances, 
one should rescale  observed limits to the actual ones upon accounting for such effects (as we have done here) or else attempt  deploying new ones adopting different selection strategies which minimise (in the case of exclusion) or indeed exalt (in the case of discovery) such effects. The reward of this approach, albeit being model-dependent, can be tantalising. On the one hand, more sensitivity can be gained to XQs and DM. On the other hand, the spin nature of the latter can be accessed. This is generally more possible in the case of XQ decays to light quarks (first and second generation) than to heavy ones (third generation).

\vspace{-2mm}


\bibliographystyle{JHEP}
\bibliography{XQCAT}

\end{document}